\documentclass[jkps,twocolumn,showkeys]{revtex4}
\usepackage[pdftex]{graphicx}
\usepackage{amssymb}
\usepackage{amsmath}
\usepackage{bm}

\begin{document}
\title[]{Scanning tunneling microscopy study of hidden phases in atomically thin $1T$-TaS$_2$}

\author{Wooin \surname{Yang}}
\author{Dowook \surname{Kim}}
\author{Hyoung Kug \surname{Kim}}
\author{Tae-Hwan \surname{Kim}}%
\email{taehwan@postech.ac.kr}
\affiliation{Department of Physics, Pohang University of Science and Technology (POSTECH), Pohang 37673, South Korea}

\date[]{Received }

\begin{abstract}
Lower thermal stability due to thinning often leads to unprecedented hidden phases in low-dimensional materials.
Such hidden phases can coexist or compete with preexisting electronic phases.
We investigate hidden phases observed in atomically thin (6--8 layers) $1T$-TaS$_2$ with scanning tunneling microscopy.
First, we can electrically induce a hidden stripe phase at room temperature.
Such a uniaxial stripe phase has three equivalent orientations by breaking three-fold symmetry of $1T$-TaS$_2$.
We also reveal that the hidden stripe phase coexists with nearly commensurate charge-density-wave phase.
Next, we observe that the emergent stripe phase spontaneously appears without any electric excitation on a tiny flake ($160\times80$~nm$^2$).
Our findings may provide a plausible explanation for the previously observed phase transition and two-fold optical response in thin $1T$-TaS$_2$ devices at room temperature.
Furthermore, the hidden stripe phase would be crucial to understand exotic CDW-related phenomena in $1T$-TaS$_2$ for potential applications. 
\end{abstract}



\pagenumbering{roman} 

\maketitle

\clearpage
\pagenumbering{arabic} 

\section{Introduction}

Macroscopic coherent states in condensed matter systems such as charge density waves (CDWs), excitonic insulators, and superconductors have often exhibited rapid transitions between the many-body ground state and a hidden state~\cite{Stojchevska2014}.
Such a hidden state typically comes back to the ground state within short time periods ($10^{-9}$--$10^{-3}$~s)~\cite{Yu1992, Cavalleri2001}.
Due to the short lifetime, it is very difficult to study hidden states by using scanning tunneling microscopy (STM), which can resolve the atomic-scale structure of systems with a relatively long imaging time ($>10$~s per image). 

Recently, $1T$-TaS$_2$ has attracted a lot of attention for its unique CDW phases and hidden states~\cite{Sipos2008, Zhao2017, Bu2019, Ravnik2019, Ma2016, Cho2016, Gerasimenko2019a, Vodeb2019}.
A high-temperature pristine triangular lattice of $1T$-TaS$_2$ starts to show quasi-periodic lattice distortion below 550~K, which leads to an incommensurate CDW (ICCDW) phase.
At $\sim 350$~K, the ICCDW phase transforms into a nearly commensurate CDW (NCCDW) phase where commensurate CDW domains are quasi-periodically separated by incommensurate domain walls~\cite{Nakanishi1977}.
With further lowering temperature, a commensurate CDW (CCDW) phase emerges below 180~K. 
In the CCDW phase or CDW domains in the NCCDW phase, 13 Ta atoms form a polaron cluster while surrounding 12 Ta atoms slightly move toward a center Ta atom, resulting in a $\sqrt{13} \times \sqrt{13} R13.9\,^{\circ}$ superstructure.
Only upon heating up, the CCDW phase transforms into a so-called triclinic phase (T-phase) with striped CCDW domains (220~K $<T<$ 282~K)~\cite{Thomson1994}. 

These complex CDW phases in $1T$-TaS$_2$ strongly suggest its metastability, which indeed allows hidden CDW phases by applying external stimuli such as pressure~\cite{Sipos2008}, strain~\cite{Zhao2017, Bu2019, Ravnik2019}, and injected holes~\cite{Ma2016, Cho2016, Gerasimenko2019a, Vodeb2019}.
In addition, efforts to tailor CDW structures have been intrigued by CDW-correlated rich quantum phases in $1T$-TaS$_2$ such as superconductivity~\cite{Sipos2008}, quantum spin liquid~\cite{Law2017}, and Mott insulating states~\cite{Qiao2017, Gerasimenko2019, Zhu2019, Butler2020}.
Especially, controlling a Mott insulating phase has been focused due to its potential application toward CDW phase based memory devices~\cite{Ma2016, Cho2016, Mraz2021}.
Furthermore, a discovery of the enhanced metastability by reducing thickness provides an extended opportunity to study hidden phases of $1T$-TaS$_2$~\cite{Yoshida2014, Yoshida2015, Patel2020}.
Very recently, a number of device applications have been demonstrated by exploiting the resistance switching around room temperature~\cite{Liu2016, Vaskivskyi2016, Mohammadzadeh2021, Taheri2022}.
While such resistance switching has been known to be originated due to the NCCDW-ICCDW transition, there has been no real-space observation to confirm existing CDW or hidden phases. 

Here, we have directly investigated hidden phases of atomically thin $1T$-TaS$_2$ by STM at room temperature. 
After electric excitation, we observe unexpected emergent stripe patterns coexisting with the expected NCCDW phase at room temperature. 
Moreover, we show that such stripe phases can exist without any electric excitation on a tiny flake (160$\times$80~nm$^2$).
Our observation indicates that the stripe phase may be a new hidden state with extremely long lifetime due to the small physical dimensions.
Further studies to control the stripe phase will deepen our understanding of hidden states of $1T$-TaS$_2$, which are fundamental for future device applications.

\section{Experimental Details}
In order to investigate hidden phases in atomically thin $1T$-TaS$_2$ with STM, we first need clean and thin enough $1T$-TaS$_2$ flakes on a conducting substrate.
To meet this challenging requirement, we prepared thin flakes via conventional mechanical exfoliation and then transferred them onto graphene epitaxially grown on a $4H$-SiC(0001) single crystalline substrate.
Detailed preparation will be found elsewhere~\cite{Kim2022}.

We repeatedly exfoliated a commercially available $1T$-TaS$_2$ crystal (2D Semiconductors Inc.) using a residue-free tape (Nitto denko Inc.) until flakes are thinned down to $<10$~layers (L). 
Although thinner flakes generally become smaller in lateral size,
we can often obtain thin $1T$-TaS$_2$ flakes with a typical lateral size of 1~$\mu$m or larger. 
All exfoliation and transfer procedures were performed in an Ar-filled glove box with O$_2$ and H$_2$O content kept below 0.1~ppm to avoid surface contamination and/or oxidation due to air exposure~\cite{Yamaguchi1997, Shen2020, Kim2022}.
We investigated the flake distribution by an optical microscope (OM) to locate where thin enough flakes were.
We selected some locations with high density of thin flakes for STM measurements.

\begin{figure}[hb]
\begin{center}
\includegraphics[width=7cm]{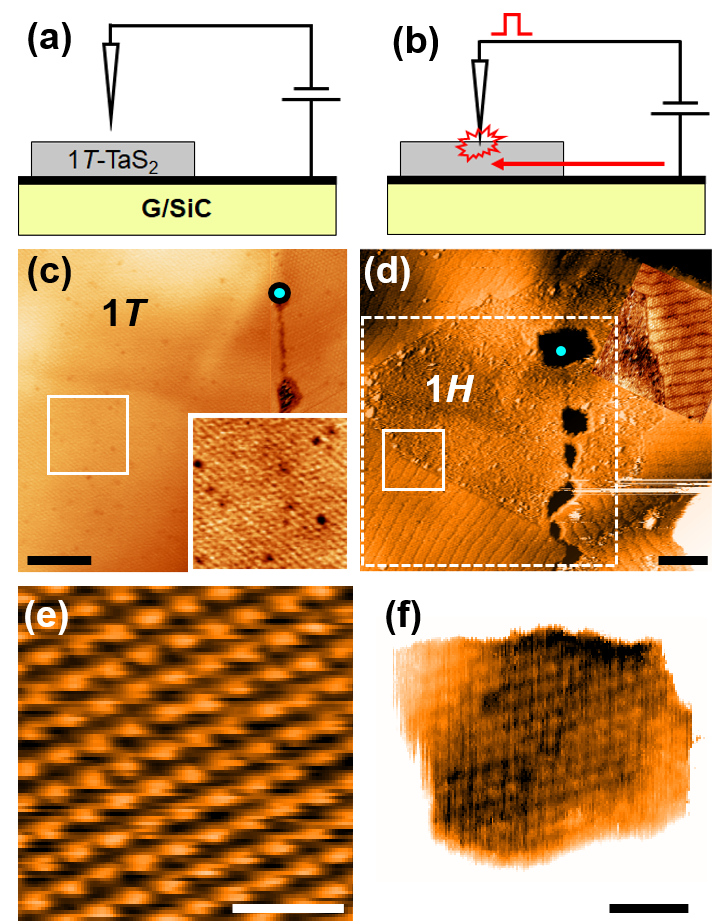}
\caption{
\textbf{a} Schematic of STM measurement. 
A bias voltage is applied to graphene epitaxially grown on $4H$-SiC(0001) with respect to the STM tip.
\textbf{b} Schematic of electric excitation.
Electric current pulses are applied through a contact between the STM tip and $1T$-TaS$_2$.
\textbf{c} STM image showing the NCCDW phase on pristine 8~L $1T$-TaS$_2$ before electric excitation. 
The inset shows a zoom-in image of the area marked with a white square. 
\textbf{d} STM image of 8~L $1T$-TaS$_2$ after the electric excitation. 
The solid white squares in \textbf{c} and \textbf{d} indicate the same area while a dashed white square shows the same area as \textbf{c}. 
Cyan circles in \textbf{c} and \textbf{d} denote the position where the electric excitation was applied.
Stripe patterns with three different orientations emerged outside the $1H$ patch (see the text).
A zoom-in image is superimposed to enhance the stripe pattern along the horizontal direction.
\textbf{e} Atom-resolved STM image exhibiting the $1\times1$ atomic structure on the triangular $1H$ patch of \textbf{d}. 
Sharp interfaces without CDW indicate that the triangular patch is transformed into the $1H$ polytype.
\textbf{f} STM image inside the monolayer deep hole at the excitation spot.
The observed CDW phase suggests that the polytype transition occurs at only the topmost layer.
Imaging conditions: 
\textbf{c} $V_{\rm b}$ = 50~mV, $I_{\rm t}$ = 100~pA; 
\textbf{d} $V_{\rm b}$ = 1~V, $I_{\rm t}$ = 20~pA; 
\textbf{e} $V_{\rm b}$ = 1~V, $I_{\rm t}$ = 100~pA; 
\textbf{f} $V_{\rm b}$ = 1~V, $I_{\rm t}$ = 20~pA. 
Scale bars: \textbf{c,d} 20~nm; \textbf{e} 1~nm; \textbf{f} 5~nm.
}
\end{center}
\end{figure}

Then, the sample was transferred into an UHV (ultrahigh vacuum) STM chamber ($P < 1 \times 10^{-8}$~Pa) using a home-built suitcase without exposure to air~\cite{park2021a, Kim2022}.
We brought an electro-chemically etched W tip around one of the target locations using a long-working-distance OM in our STM setup~\cite{Kim2022}.
Figure~1a shows the typical electric configuration between the sample and STM tip. 
All STM topographic images were recorded in a constant-current mode at room temperature where bulk $1T$-TaS$_2$ shows the NCCDW phase.

For electric excitation, we placed the tip over a specific spot on the sample. 
While the STM feedback was shut off, the sample bias was set to 10~mV.
Sequentially, the tip was moved toward the sample surface to make electric contact between the tip and sample.
Then, the sample bias pulse (1~V, 1~ms) was applied to introduce electric excitation as shown in Fig.~1b.
After the electric excitation, the tip was moved back away from the surface, the original sample bias was set back, and the STM feedback was on.

\section{Results and Discussion}
Before any electric excitation, we can observe quasi-hexagonal commensurate CDW domain structures separated by incommensurate domain walls on thin $1T$-TaS$_2$ (Fig.~1c), which is expected in the NCCDW phase at room temperature~\cite{Yang2022a}.
Fast Fourier transform (FFT) of atom-resolved STM topographic images provides detailed information of CDW domain structures on thin flakes, which enables direct comparison between thin and bulk $1T$-TaS$_2$. 
From the FFT analysis on the 8~L flake, we found that the average CDW periodicity (1.17~nm) and tilted CDW angle (12.1$^\circ$) with respect to the atomic lattice are quite similar to the bulk values (1.18~nm and 11.8$^\circ$).
On the other hand, the smaller size of commensurate CDW domains on thin flakes is attributed to lower thermal stability due to thinning~\cite{Yang2022a}.
Even though thinner flakes show smaller CDW domains, the pristine CDW phase can be considered as the same ground state as bulk.

As previously demonstrated, atomically thin $1T$-TaS$_2$ exhibits the enhanced metastability in contrast to the bulk counterpart~\cite{Yang2022a}.
In order to induce any hidden phases at room temperature, we employ electric excitation similar to what has been done in $1T$-TaS$_2$ devices~\cite{Liu2016, Vaskivskyi2016, Mohammadzadeh2021, Taheri2022}.
As described above, we placed the STM tip over a specific spot (Fig.~1c) and introduced an electric excitation.
Later, we notice significant changes due to the electric excitation as shown in Fig.~1d.
First, surface etching happens at the spot, showing that only the topmost layer was removed.
Second, CDW domains disappear on a triangular patch surrounding the spot.
Atom-resolved STM images reveal that a polytype transformation from $1T$ to $1H$ occurs on the triangular patch (Fig.~1e).
Last, we found stripe structures with three different orientations outside the triangular patch, indicating that symmetry breaking due to the electric excitation leads to three energetically equivalent orientations.

Since TaS$_2$ has the smallest energy difference between the $1T$ and $2H$ polymorphs among known TMDs, a $T$-to-$H$ polytype transformation can be induced on the topmost monolayer of bulk $1T$-TaS$_2$ by electric pulses~\cite{Kim1997, Ma2016}, optical pulses~\cite{Ravnik2019, Ravnik2021}, or vacuum thermal annealing~\cite{Wang2018}.
Especially, vacuum thermal annealing induces the monolayer surface etching as well as the $T$-$H$ polytype transformation at the same time as shown in Fig.~1d.
In this sense, the electric excitation can be considered to generate enough local Joule heating for the observed local surface etching and the polytype transformation on thin $1T$-TaS$_2$. 
This consideration is confirmed by a 1$\times$1 atomic lattice without CDW (Fig.~1e) and atomically well defined boundaries between $1T$ and $1H$ structures, which prefer to align along the atomic lattice~\cite{Kim1997}.
Similarly to the annealed bulk $1T$-TaS$_2$~\cite{Wang2018}, CDW-resolved STM images within the pit reveal that the bottom layer preserves the preexisting NCCDW phase of $1T$-TaS$_2$ (Fig.~1f).
Thus, our result indicates that thin $1T$-TaS$_2$ at room temperature can experience the $T$-to-$H$ polytype transformation confined on the topmost layer due to the local Joule heating.

\begin{figure}[tb]
\begin{center}
\includegraphics[width=7cm]{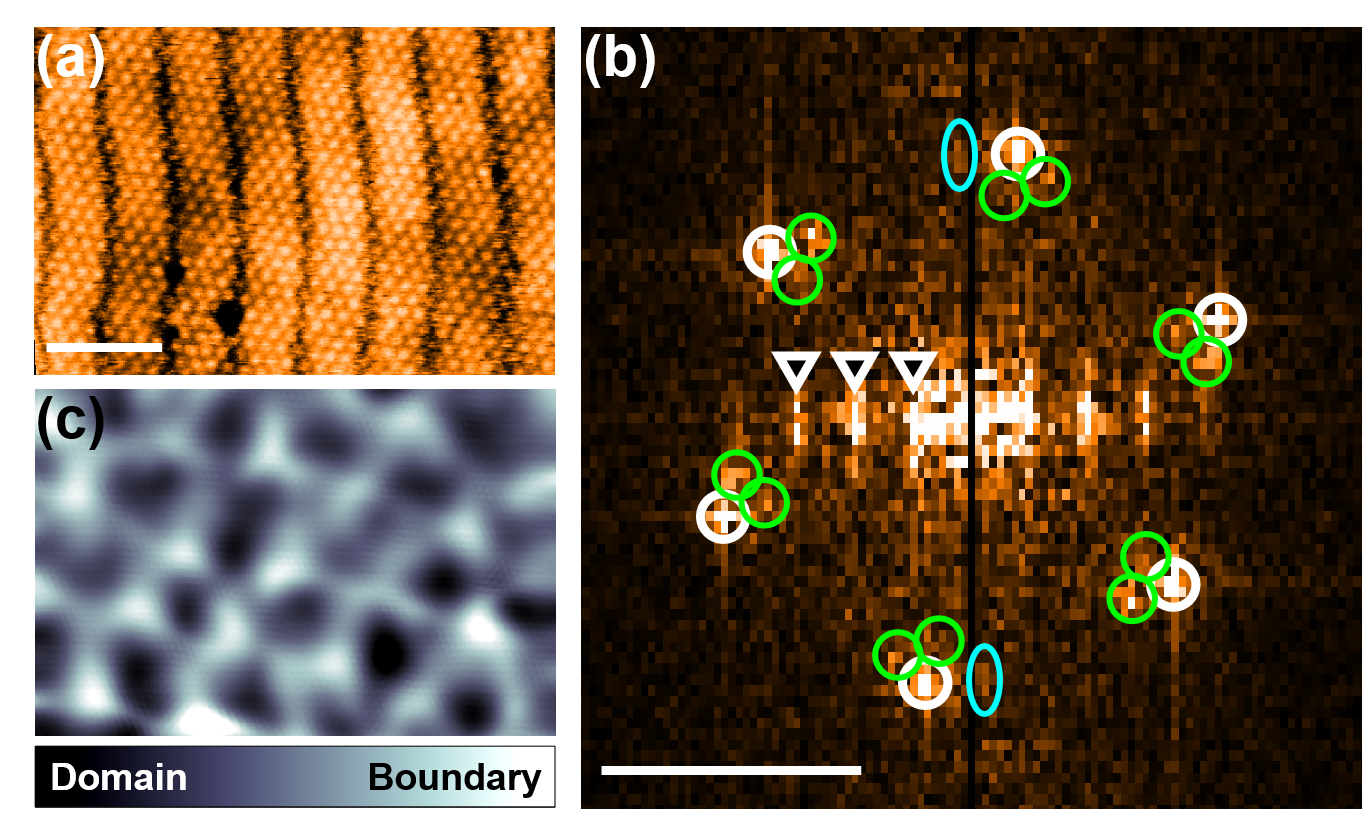}
\caption{
Coexisting stripe and CDW domains.
\textbf{a} STM topographic image of coexisting stripe and CDW.
Imaging condition: $V_{\rm b}$ = 1~V, $I_{\rm t}$ = 100~pA. 
\textbf{b} Fast Fourier Transform (FFT) of \textbf{a}. 
FFT peaks from the stripe phase are denoted by white triangles while CDW peaks and their satellite peaks are denoted by white and green circles, respectively. 
A pair of cyan ellipses represents satellite peaks due to the stripe phase. 
\textbf{c} CDW orientation map of \textbf{a}. 
Contrasts in the orientation map represent CDW phase shifts that become larger at domain boundaries.
Scale bar: \textbf{a} 10~nm, \textbf{b} 5~nm$^{-1}$.
}
\end{center}
\end{figure}

In addition to the polytype transition, we can observe the unexpected stripe phase which has not been reported in bulk $1T$-TaS$_2$.
More surprisingly, the stripe phase coexists with the pristine CDW domains (Fig.~2).
In sharp contrast to the spatially confined polytype transition, the stripe phase is observed over the whole flake with a lateral size of 1200$\times$400~nm$^2$. 
This global emergence strongly suggests that the stripe phase is driven mainly by dissipated electric current rather than the localized Joule heating.
Due to the uniaxial nature, one may consider such a stripe as a T-phase in bulk even though the T-phase exists only during warming up bulk $1T$-TaS$_2$ from the CCDW phase~\cite{Thomson1994}.
However, CDW-resolved STM images provide compelling evidence that this stripe phase is not the same as the T-phase observed in bulk $1T$-TaS$_2$.
From the atomic boundaries between the $T$-type and $H$-type, we unambiguously determine that the observed stripes are roughly perpendicular to the atomic axes, clearly distinct from the reported T-phase ($10\,^{\circ}$ off the atomic axes)~\cite{Burk1992}.

We could obtain STM topographic images of the stripe phase coexisting with CDW domains (Fig.~2a).
The FFT image of Fig.~2a clearly shows the coexistence of two different phases as shown in Fig.~2b, indicating the almost intact NCCDW phase even with the stripe phase.
The FFT peaks and their higher harmonics originated from the stripe phase appear along the horizontal axis.
Although most of satellite peaks of CDW domains indicate the persistent NCCDW phase, 
some weak pairs of the satellite peaks suggest a possibility for interrupting CDW domains due to the stripe phase.
In order to check such a possibility, we visualize the CDW domain structure by carefully filtering the stripe phase from the original STM image and tracking domain boundaries along three different orientations (see more details in Appendix)~\cite{Gerasimenko2019a, Altvater2021} (Fig.~2c). 
From Fig.~2c, we conclude that the NCCDW phase coexists with the stripe phase at room temperature while a few NCCDW domain walls are slightly suppressed due to the superimposed stripe phase, suggesting two different phases do not interact significantly. 
Thus, this apparently independent stripe phase is completely distinct from any known CDW phases including the T-phase of bulk TaS$_2$.

\begin{figure}[tb]
\begin{center}
\includegraphics[width=7cm]{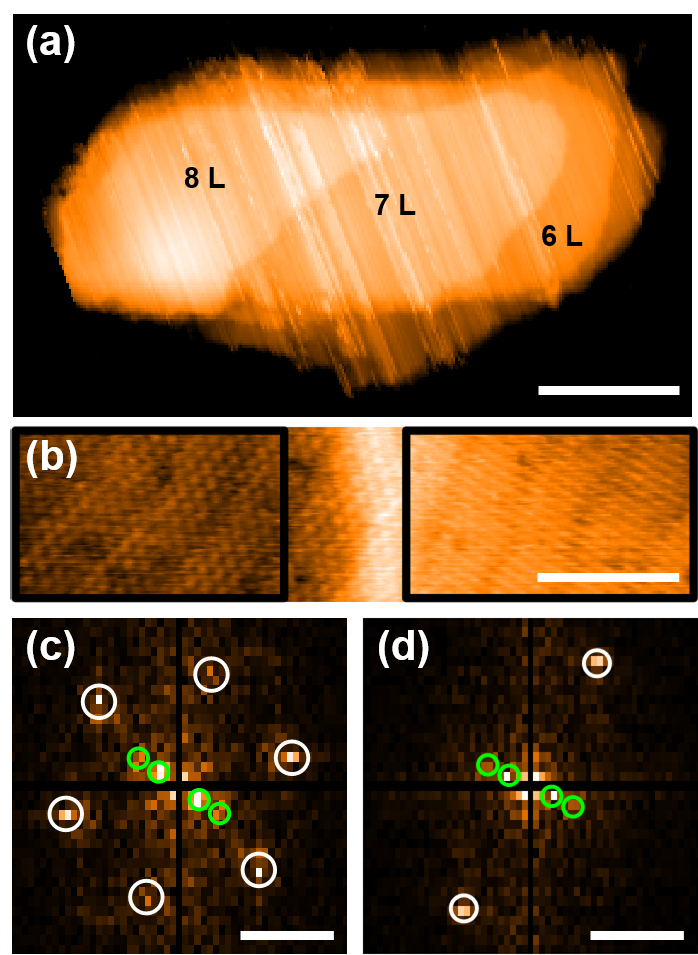}
\caption{
Spontaneously emerging stripe phase on a tiny flake without electric excitation.
\textbf{a} STM image of a thin $1T$-TaS$_2$ with a thickness of 6--8~L and a small lateral size of 160$\times80$~nm$^2$. 
\textbf{b} Zoom-in STM image showing coexisting NCCDW and stripe phases on the 7~L $1T$-TaS$_2$. 
\textbf{c},\textbf{d} FFT images of the left and right sides of \textbf{b} indicated by rectangles. 
The left side shows the same stripe phase (green circles) coexisting with the NCCDW phase (white circles).
On the other hand, the right side exhibits only the stripe phase (green circles) without any CDW polaron cluster.
Instead of CDW clusters, the $3\times1$ atomic structure (white circles) appears on the right side.
The affine transform was applied for quantitative analysis. 
CDW peaks were used as a reference for the affine transform.
Imaging conditions: 
\textbf{a} $V_{\rm b}$ = 2~V, $I_{\rm t}$ = 50~pA; 
\textbf{b} $V_{\rm b}$ = 0.4~V, $I_{\rm t}$ = 100~pA.
Scale bars: \textbf{a} 40~nm; \textbf{b} 10~nm; \textbf{c},\textbf{d} 5~nm$^{-1}$.
}
\end{center}
\end{figure}

We occasionally observe very tiny flakes as shown in Fig.~3a.
The flake has a thickness of 6--8~L and a lateral size of $160\times80$~nm$^2$.
Most strikingly, on such a tiny flake, we could observe the same coexistence of the NCCDW and stripe phases even without any electric excitation (left of Fig.~3b).
At the same time, we also observe that the NCCDW phase completely disappears but the stripe phase remains on the same layer (right of Fig.~3b).
From the FFT analysis, we confirm that the NCCDW and stripe phase still coexist on one location (Fig.~3c) but the stripe survives on the other location (Fig.~3d).
Furthermore, we notice the $3\times1$ atomic structure on the stripe phase without the NCCDW phase along one of the atomic axes (Fig.~3b and d).
Such a $3\times1$ structure has not been reported in bulk $1T$-TaS$_2$.
Since the $3\times3$ CDW phase is observed in another polytype $2H$-TaS$_2$~\cite{Slough1986}, one may consider that the $3\times1$ structure would be originated from the $H$ structure.
However, we cannot expect any CDW formation at room temperature because the CDW transition occurs below 75~K.
In addition, the boundary between two distinct regions does not show any abrupt transition like the $T$/$H$ interface in Fig.~1d.
Thus we can safely rule out the possibility that the $3\times1$ structure would result from the $H$ polytype.
This observation suggests that the smaller lateral size significantly enhances the metastability of CDW phases even further and the emergent phase behaves somewhat differently from one induced by the electric excitation.

\begin{figure}[tb]
\begin{center}
\includegraphics[width=7cm]{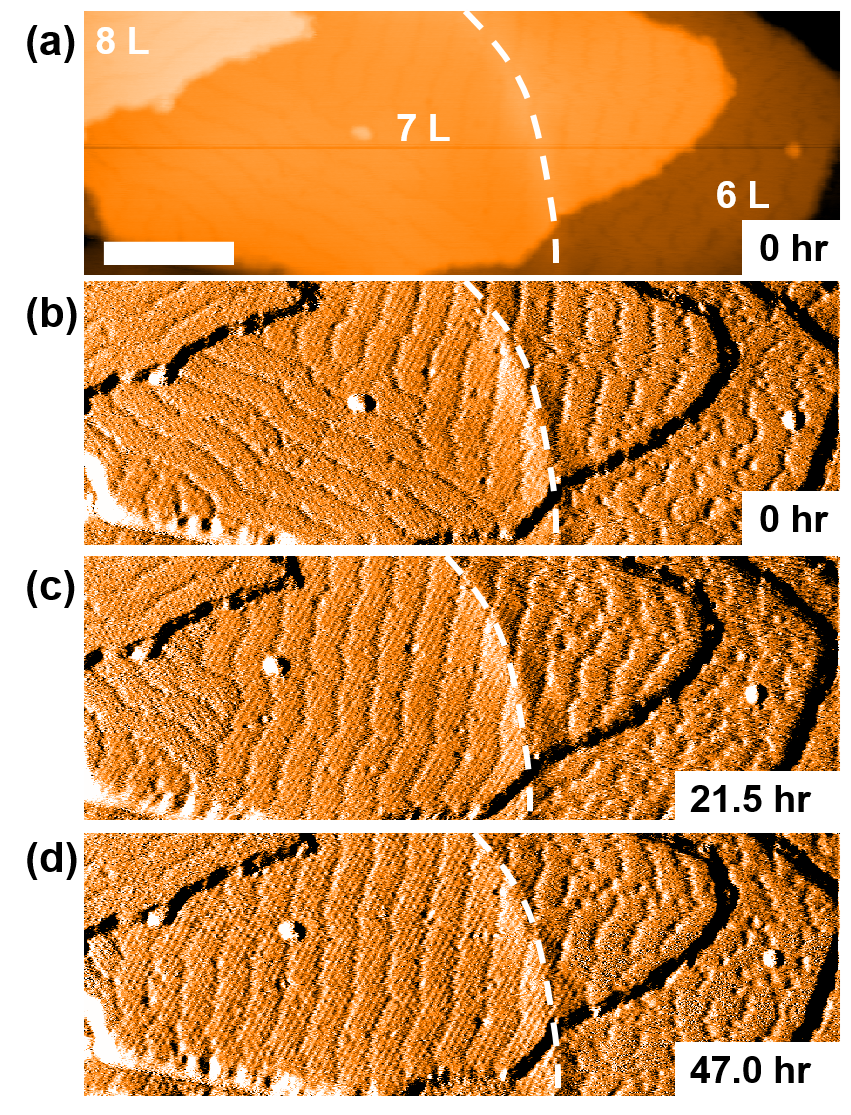}
\caption{
Time evolution of competing stripe phases in a tiny flake.
\textbf{a} STM topographic image of 6--8~L.
The white dashed line indicates a phase boundary separating two regions: (left) the coexisting NCCDW and stripe phase and (right) the stripe phase with the $3\times1$ atomic structure. 
Scale bar: 20~nm.
\textbf{b} Derivative STM image of \textbf{a} enhancing the fine structures. 
The stripe phase exists over the whole flake. 
\textbf{b--d} Series of STM images showing the same area as a function of time.
On the 7~L $1T$-TaS$_2$, one stripe phase gradually grows at the expense of the other stripe phases with different orientations with increasing time.
Imaging conditions: 
\textbf{a},\textbf{b} $V_{\rm b}$  = 1~V, $I_{\rm t}$ = 100~pA; 
\textbf{c} $V_{\rm b}$  = 1~V, $I_{\rm t}$ = 130~pA; 
\textbf{d} $V_{\rm b}$  = 1~V, $I_{\rm t}$ = 70~pA.
}
\end{center}
\end{figure}

Our finding can be corroborated by following repeated observations on the tiny flake.
We repeatedly scan over the whole flake to see any phase fluctuation with time as shown in Fig.~4.
With increasing time, we observe that one of the stripe phases gradually grows at the expense of the other two stripe phases with different orientations on 7~L while the configuration of the stripe phase remains on 8~L.
The dashed line indicates the boundary between two distinct regions described in Fig.~3.
As demonstrated in Fig.~4b, thinner layers seem to prefer the stripe phase rather than the NCCDW phase.
Although characteristic stripe features are observed over 6--8~L, thinner layers show more noisy and less straight stripe patterns with increasing time. 
After 47~hours, the $3\times1$ structure still exists on 6~L while the stripe phase almost loses its uniaxial preference.
This finding strongly suggests that the NCCDW phase can stabilize the stripe phase.
These thickness- and time-dependent stripe phase indicates the enhanced metastability of $1T$-TaS$_2$ on the tiny flake, which can lead to more diverse CDW-related device applications.

One may consider the substrate effect to explain the origin of the hidden phases.
However, recent STM experiment on graphene covered $1T$-TaS$_2$ revealed the intact NCCDW phase of TaS$_2$~\cite{Altvater2021a}, which strongly suggests that graphene rarely interacts with TaS$_2$.
In this sense, our epitaxial graphene would not significantly interact with thin TaS$_2$ flakes.
Thus, we can exclude the substrate effect causing the observed hidden phase.

Our finding would provide a new insight into understanding the phase transition in thin $1T$-TaS$_2$ devices.
In general, it has been accepted that the NCCDW-ICCDW phase transition occurs by external stimulus at room temperature. 
However, we demonstrate the hidden phase can be created after electric excitation.
Thus, we should systematically investigate the phase transition in thin $1T$-TaS$_2$ devices due to electric excitation by measuring with local and global probes at the same time.

On the other hand, our finding clearly demonstrates that the uniaxial stripe hidden phases can support the symmetry lowering observed in optically excited $1T$-TaS$_2$~\cite{Lacinska2022}. 
Without the hidden phases, the recent 2-fold optical activity in optically excited $1T$-TaS$_2$ cannot be explained with the 3-fold rotational symmetry of pristine $1T$-TaS$_2$. 
In particular, one stripe phase grows at the expense of two other competing stripe phases (Fig.~4).
Such a time evolution toward the single orientation strongly suggests that $1T$-TaS$_2$ prefers to have lower symmetry at a given environment under excitation.

\section{Summary}

We investigated the emergent stripe phase of atomically thin $1T$-TaS$_2$ by STM at room temperature. 
The stripe phase was induced by electric excitation between the STM tip and thin TaS$_2$ flake.
We found the local polymorphic transformation due to Joule heating as well as the coexistence of the emergent stripe and NCCDW phases after electric excitation.  
We observed the same stripe phase on a tiny flake without electric excitation due to its small lateral size.
The thickness- and time-dependent phase fluctuation in the stripe phase suggests the enhanced metastability of CDW phases on the tiny $1T$-TaS$_2$ flake.
Our finding demands further studies to control such a stripe phase and to enrich our understanding on CDW phases of thin $1T$-TaS$_2$, which is crucial for CDW-related device applications. 

\begin{figure*}[ht]
\includegraphics[width=\textwidth]{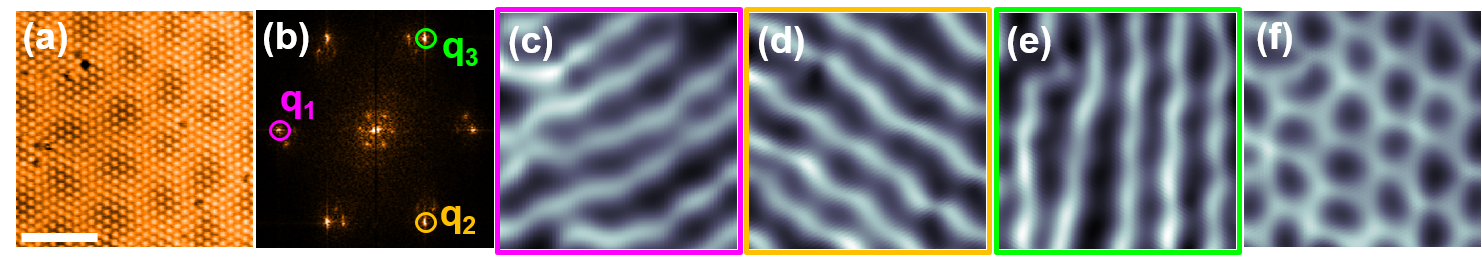} 
\caption{
\textbf{a} STM topograph on 11~L with distinct CDW domain contrast ($V_{\rm b}$  = 1~V, $I_{\rm t}$ = 100~pA, Scale bar: 10~nm).
\textbf{b} FFT of \textbf{a}. 
Each $\vec{q}_i$ indicates the corresponding CDW peak (denoted by a circle). 
\textbf{c--e} Local phase gradient maps along $\vec{q}_1$, $\vec{q}_2$, and $\vec{q}_3$, respectively.
\textbf{f} CDW orientation map by merging \textbf{c}--\textbf{e}. 
}
\label{r3}
\end{figure*}

\section{appendix}

A CDW modulation with a characteristic wave vector $\vec{q}_i (i=1,2,3)$ can be expressed as the plane wave with local phase $\delta(\vec{r})$: $A\, \cos(\vec{q_i} \cdot \vec{r} + \delta(\vec{r}))$,
where $A$ and $\vec{r}$ denote the CDW amplitude and position vector, respectively.
The CDW modulation has the almost same phase within commensurate CDW domains while neighboring CDW domains exhibit different phases.
If we extract such phase differences between neighboring CDW domains, we can visualize domain walls as follows.
We first obtain the fast Fourier transform image to identify the $\vec{q}_i$ vectors (Fig.~5b).
Next, we use the combined plane-fit and gradient algorithm~\cite{Yang2007} to get local phase gradient maps (Fig.~5c--e).
Then, we can get the CDW orientation map by merging these gradient maps along three different $\vec{q}_i$ directions (Fig.~5f).

\begin{acknowledgments}
This work was supported by the National Research Foundation of Korea (NRF) funded by the Ministry of Science and ICT, South Korea (Grants No. NRF-2017R1A2B4007742, 2021R1F1A1063263, 2021R1A6A1A10042944, and 2022M3H4A1A04074153).
\end{acknowledgments}


\end{document}